\documentclass[%
reprint,
superscriptaddress,
 amsmath,amssymb,
 aps,
prb,
]{revtex4-1}

\usepackage{graphicx}
\usepackage{dcolumn}
\usepackage{bm}
\usepackage{subfigure}
\usepackage{dcolumn} 

\begin{document}


\title{Novel phase of beryllium fluoride at high pressure}

\author{Maksim S. Rakitin}
\email{Correspondence and requests for materials should be addressed to
M.S. Rakitin (maksim.rakitin@stonybrook.edu) or A.R. Oganov
(artem.oganov@stonybrook.edu)}
\affiliation{Department of Geosciences, State University of New York,
Stony Brook, NY 11794, USA}

\author{Artem R. Oganov}
\affiliation{Skolkovo Institute of Science and Technology, Skolkovo Innovation
Center, Bldg.~3, Moscow 143026, Russia}
\affiliation{Department of Geosciences, State University of New York, Stony
Brook, NY 11794, USA}
\affiliation{Moscow Institution of Physics and Technology, 9 Institutskiy Lane,
Dolgoprudny City, Moscow Region 141700, Russia}
\affiliation{School of Materials Science, Northwestern Polytechnical University,
Xi'an 710072, China}

\author{Haiyang Niu}
\affiliation{Moscow Institution of Physics and Technology, 9 Institutskiy Lane,
Dolgoprudny City, Moscow Region 141700, Russia}
\affiliation{Shenyang National Laboratory for Materials Science, Institute of
Metal Research, Chinese Academy of Sciences, Shenyang 110016, China}

\author{M. Mahdi Davari Esfahani}
\author{Xiang-Feng Zhou}
\author{Guang-Rui Qian}
\affiliation{Department of Geosciences, State University of New York,
Stony Brook, NY 11794, USA}

\author{Vladimir L. Solozhenko}
\affiliation{LSPM-CNRS, Universit{\'e} Paris Nord, 93430 Villetaneuse, France}

\date{\today}

\begin{abstract}
A previously unknown thermodynamically stable high-pressure phase of BeF$_{2}$
has been predicted using the evolutionary algorithm USPEX. This phase occurs in
the pressure range 18--27~GPa. Its structure has $C2/c$ space group symmetry and
contains 18 atoms in the primitive unit cell. Given the analogy between
BeF$_{2}$ and SiO$_{2}$, silica phases have been investigated as well, but the
new phase has not been observed to be thermodynamically stable for this system.
However, it is found to be metastable and to have comparable energy to the known
metastable phases of SiO$_{2}$, suggesting a possibility of its synthesis.
\begin{description}
\item[PACS numbers]
\end{description}
\end{abstract}

\pacs{Valid PACS appear here}
\maketitle


\section{\label{sec:intro}Introduction}

\begin{figure}[hbt]
\includegraphics[width=\columnwidth]{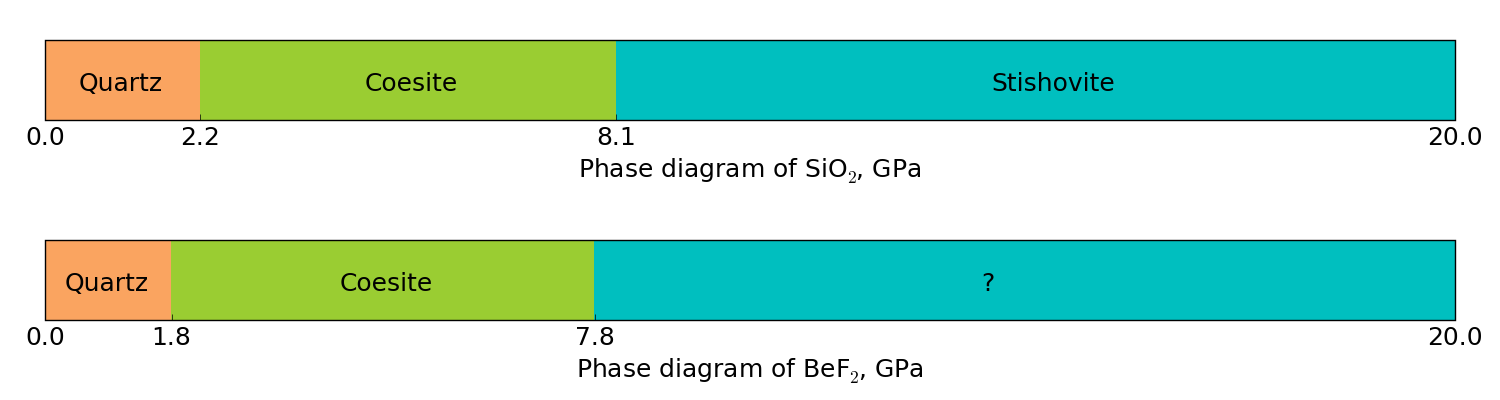}
\caption{\label{fig:exp_diagram} Phase diagrams of SiO$_{2}$ \cite{Swamy1994}
and BeF$_{2}$ \cite{Ghalsasi2011} at low (up to room) temperatures.}
\end{figure}

Beryllium fluoride has many applications, such as coolant component in molten
salt nuclear reactors \cite{Weaver1961, Benes2009}, production of special
glasses \cite{Parker1989, Gan1995}, manufacture of pure beryllium
\cite{Hausner1965}, etc. Structurally, BeF$_{2}$ phases are similar to the
phases of SiO$_{2}$ (Fig.~\ref{fig:exp_diagram}): $\alpha$-quartz phase of
BeF$_{2}$ and SiO$_{2}$ is stable from 0 to $\sim$2~GPa, and then transforms to
coesite phase which persists up to $\sim$8~GPa, and then transforms to
stishovite (rutile-type phase) in SiO$_{2}$ \cite{Swamy1994}. However, the
behavior of BeF$_{2}$ experimentally is not known for pressures above 8~GPa (see
Scheme~1 in Ref. \cite{Ghalsasi2011}).

One of our goals in present paper is to reveal which phase transitions can occur
at higher pressures in BeF$_{2}$. Beryllium compounds are extremely toxic for
humans, and this limits experimentation. Computer simulation is a safe and cheap
alternative to investigate such structures. In recent \emph{ab initio} study
\cite{Yu2013} authors explored 13 well-known AB$_{2}$ structure types for their
possible stability for BeF$_{2}$: $\alpha$-quartz-type ($P3_{1}21$),
$\beta$-quartz-type ($P6_{2}22$), $\alpha$-cristobalite-type ($P4_{1}2_{1}2$),
$\beta$-cristobalite-type ($Fd$-$3m$), cubic CaF$_{2}$-type ($Fm$-$3m$),
$\alpha$-PbCl$_{2}$-type ($Pnma$), Ni$_{2}$In-type ($P6_{3}/mmc$), coesite-type
($C2/c$), rutile-type ($P4_{2}/mnm$), baddeleyite-type ($P2_{1}/c$),
$\alpha$-PbO$_{2}$-type ($Pbcn$), $\alpha$-CaCl$_{2}$-type ($Pnnm$) and
pyrite-type ($Pa$-$3$) structures. They found that the sequence of
pressure-induced phase transitions of BeF$_{2}$ up to 50~GPa is as follows:
$\alpha$-quartz-type $\xrightarrow{0.59~GPa}$ coesite-type
$\xrightarrow{6.47~GPa}$ rutile-type $\xrightarrow{24.94~GPa}$
$\alpha$-PbO$_{2}$-type structures. Although BeF$_{2}$ under pressure has been
theoretically investigated by Yu \emph{et al.} \cite{Yu2013}, we revisit these
results to check for previously unknown structure(s), and we explore the
relevance of these findings for SiO$_{2}$.

\section{\label{sec:comp}Computational details}
Computer simulations of BeF$_{2}$ and SiO$_{2}$ has been performed in two steps:
(1) prediction of a new structure of BeF$_{2}$ using USPEX evolutionary
algorithm; (2) calculation of properties of BeF$_{2}$ and SiO$_{2}$ in the wide
range of pressures from 0 to 50~GPa with a 1~GPa step using DFT.

To find stable lowest-energy crystals structures, we performed fixed-composition
search of the BeF$_{2}$ system at different pressures (15, 20 and 25~GPa) using
the USPEX code \cite{Oganov2006, Oganov2011, Lyakhov2013}, in conjunction with
first-principles structure relaxations using density functional theory (DFT)
within the Perdew-Burke-Ernzerhof (PBE) generalized gradient approximation (GGA)
\cite{Perdew1996}, as implemented in the VASP package \cite{Kresse1996}. We
employed projector augmented wave (PAW) \cite{Kresse1999} potentials with 2
valence electrons for Be and 7 --- for F. The wave functions were expanded in a
plane-wave basis set with the kinetic energy cutoff of 600~eV and
$\Gamma$-centered meshes for Brillouin zone sampling with reciprocal space
resolution of 2$\pi\times$0.10~\r{A}$^{-1}$.

We used the VASP package to carefully reoptimize the obtained structures before
calculating phonons, elasticity, electronic density of states (DOS), hardness of
BeF$_{2}$ and SiO$_{2}$. For these relaxations, we also used the plane-wave
cutoff of 600~eV and \emph{k}-meshes with resolution of 0.10~\r{A}$^{-1}$.
Phonons calculations have been performed using Phonopy \cite{phonopy} and
Quantum Espresso \cite{quantum-espresso} codes for the relaxed structures at
pressures where these structures are found to be thermodynamically stable.
Hardness was calculated using 3 methods: Lyakhov-Oganov model \cite{Lyakhov2011}
based on the strength of bonds between atoms and bond network topology, Chen-Niu
model \cite{Chen2011} which uses elastic constants obtained from DFT
calculations and Mukhanov-Kurakevych-Solozhenko thermodynamic model of hardness
\cite{Mukhanov2008}.

\section{\label{sec:results}Results and discussion}

\begin{figure}[hbtp]
\includegraphics[scale=0.21]{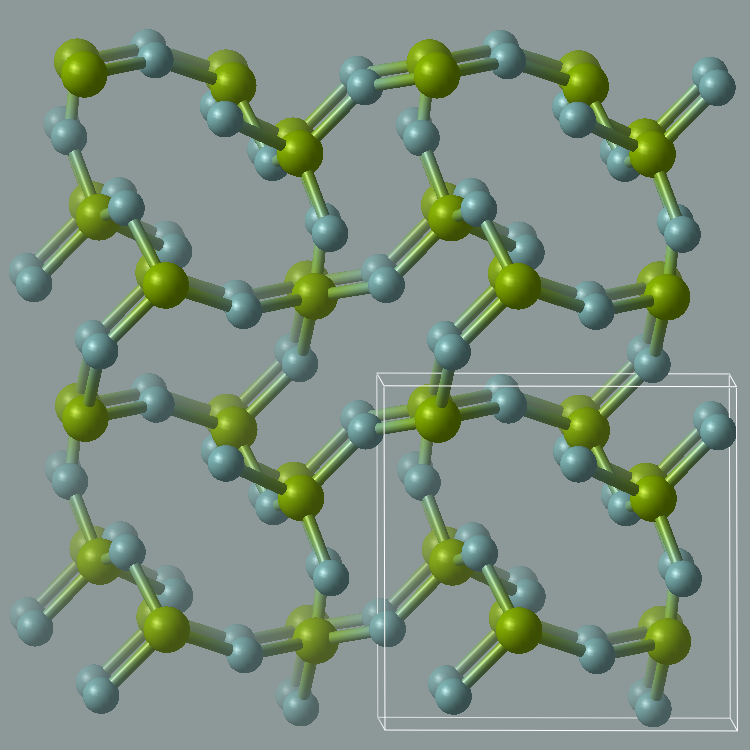}
\includegraphics[scale=0.21]{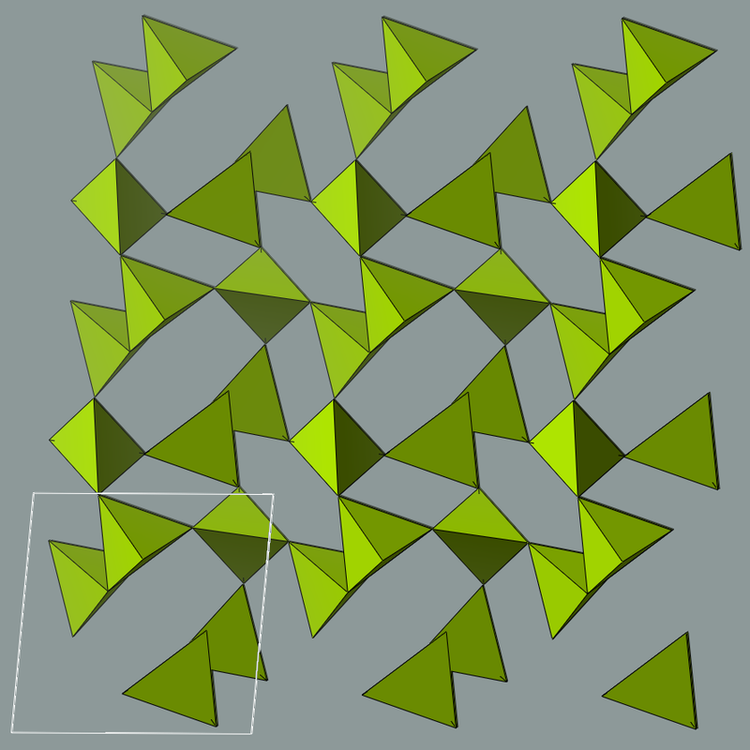}
\caption{\label{fig:new_struct}$C2/c$ structure of BeF$_{2}$, stable at
18--27~GPa.}
\end{figure}

\begin{figure}[hbtp]
\includegraphics[width=\columnwidth]{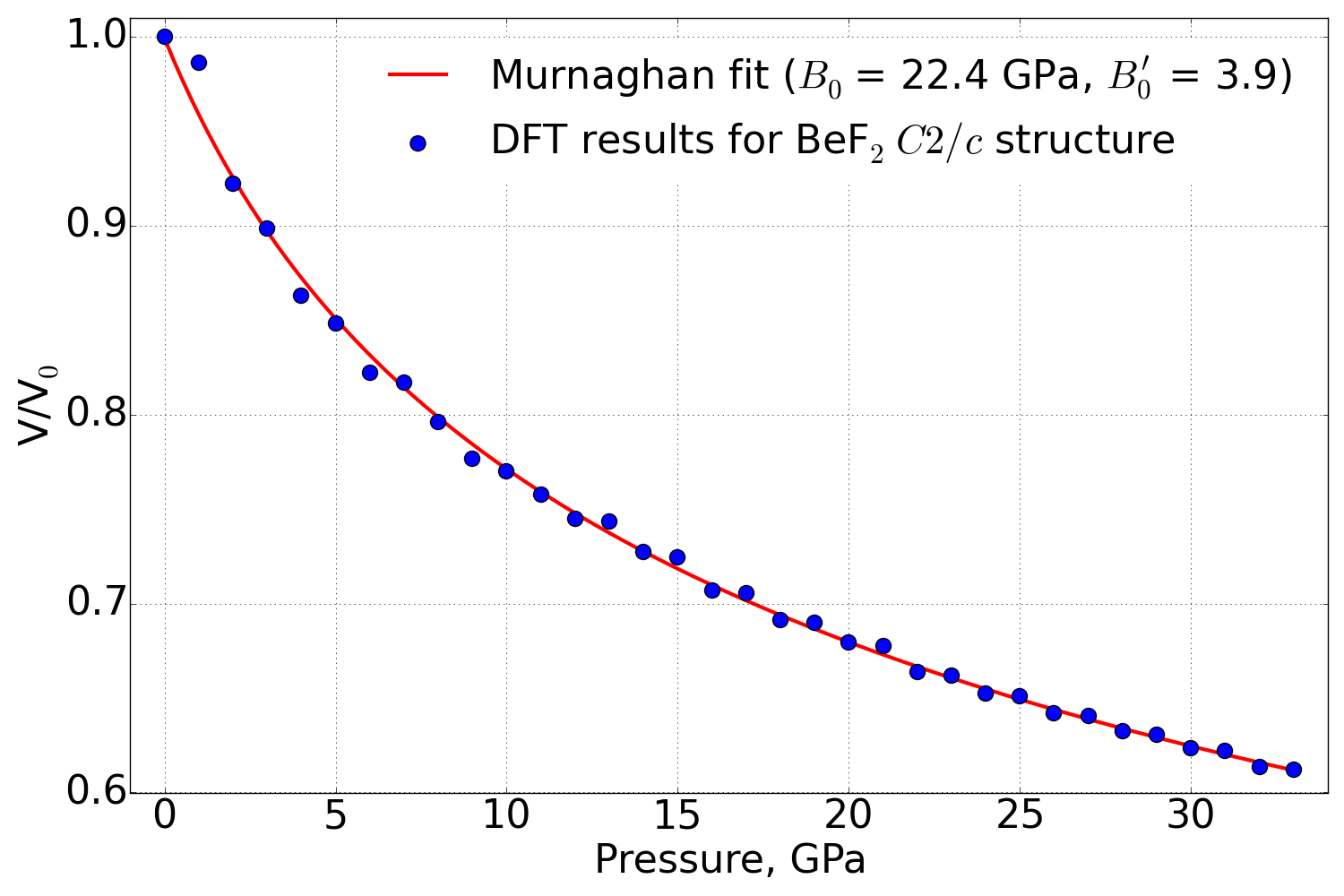}
\caption{\label{fig:BeF2_new_EOS}Equation of state of BeF$_{2}$ $C2/c$
structure.}
\end{figure}

USPEX allowed us to find a new structure of BeF$_{2}$, stable at 18--27 GPa
(Fig.~\ref{fig:new_struct}). The structure has $C2/c$ space group and contains
12 formula units in the Bravais cell (6 in the primitive cell) with
\emph{a}=8.742~\r{A}, \emph{b}=8.695~\r{A}, \emph{c}=4.178~\r{A} and
$\beta$=66.1$^{\circ}$ (at 20~GPa). Calculated density of this new $C2/c$ phase
is 4.2\% higher than density of coesite phase, both at 20~GPa. For reference,
here are lattice parameters for BeF$_{2}$-stishovite at 30~GPa:
\emph{a}=\emph{b}=3.986~\r{A}, \emph{c}=2.501~\r{A} and
$\alpha$=$\beta$=$\gamma$=$90^{\circ}$. The value of the bulk modulus
$B_{0}$=22.4~GPa of the $C2/c$ structure of BeF$_{2}$ with its pressure
derivative $B_{0}^{\prime}$=3.9 was obtained from a least-squares fit using the
Murnaghan equation of state \cite{Murnaghan1944} (Fig.~\ref{fig:BeF2_new_EOS}).
The zero-pressure unit cell volume was taken as $V_{0}$=213.7~\r{A}$^{3}$.

\subsection{Thermodynamic stability}

\begin{figure}[htbp]
\subfigure[]{
\includegraphics[width=\columnwidth]{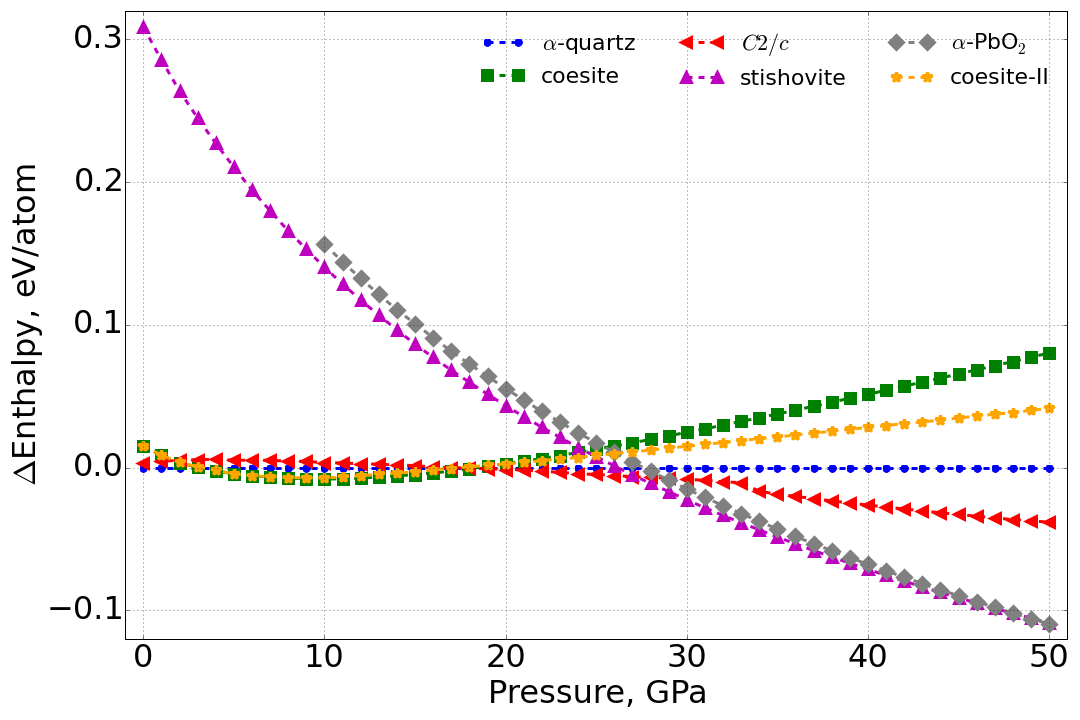}
\label{fig:enth_BeF2}
}
\subfigure[]{
\includegraphics[width=\columnwidth]{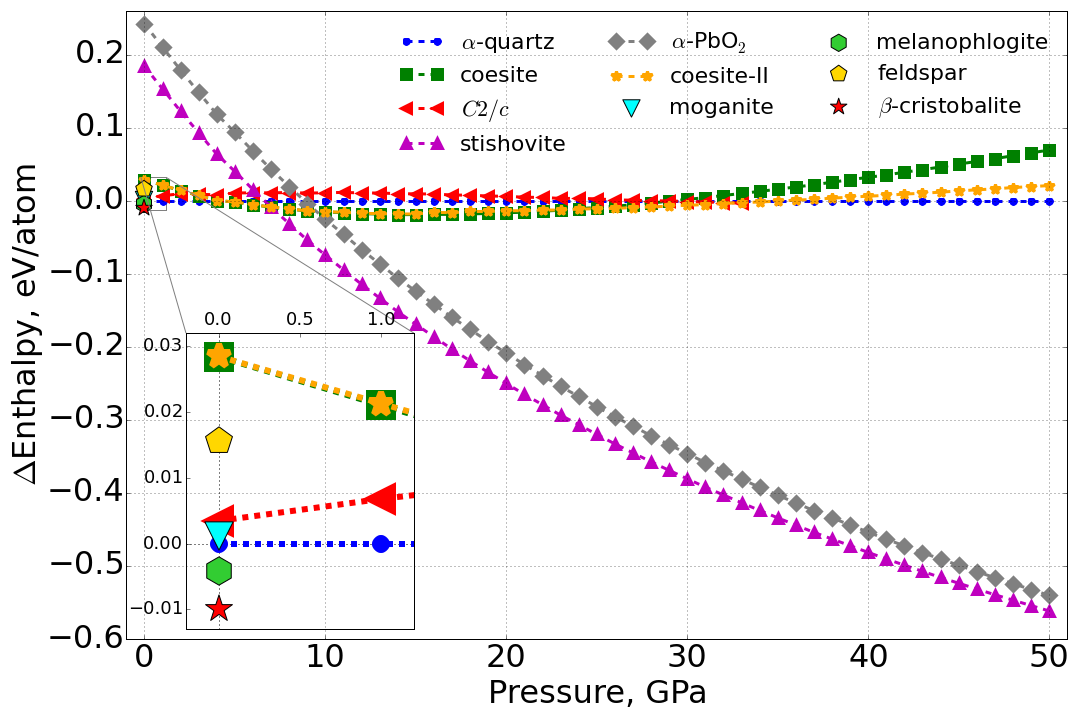}
\label{fig:enth_SiO2}
}
\caption{\label{fig:enth_SiO2_BeF2}Enthalpies (relative to $\alpha$-quartz) of
\subref{fig:enth_BeF2} BeF$_{2}$ and \subref{fig:enth_SiO2} SiO$_{2}$ phases as
a function of pressure.}
\end{figure}

We have calculated the enthalpies of $\alpha$-quartz ($P3_{2}21$), coesite
($C2/c$), coesite-II ($C2/c$), stishovite ($P4_{2}/mnm$),
$\alpha$-PbO$_{2}$-type ($Pbcn$) structure and our new structure ($C2/c$) for
both BeF$_{2}$ and SiO$_{2}$ at different pressures from 0 to 50~GPa with a
1~GPa step. The results are presented in Fig.~\ref{fig:enth_SiO2_BeF2}.

\subsubsection{BeF$_{2}$ under pressure}
For the case of BeF$_{2}$ $\alpha$-quartz structure is stable from 0 to 4~GPa,
followed by coesite structure stable from 4 to 18~GPa, and the $C2/c$ structure
is found to be stable between 18 and 27~GPa, which then gives place to
stishovite structure at higher pressures (Fig.~\ref{fig:enth_BeF2}). We see
transition from coesite-type to $C2/c$, then to rutile-type, but at much higher
pressure (27~GPa against 6.47~GPa in Ref. \cite{Yu2013}, where LDA was used).
According to Demuth \emph{et al.} \cite{Demuth1999}, the LDA approximation used
in Ref. \cite{Yu2013} underestimates phase transition pressures, whereas using
the GGA yields more reliable results. The $\alpha$-PbO$_{2}$-type structure is
not stable at any pressure (in the investigated interval from 0 to 50~GPa) for
BeF$_{2}$ (though it is close to stability at $\sim$27~GPa), while for SiO$_{2}$
it is indeed stable at pressures above $\sim$80--90~GPa \cite{Oganov2005}.

\subsubsection{SiO$_{2}$ under pressure}
From Fig.~\ref{fig:enth_SiO2} it is clearly seen that in SiO$_{2}$ the
transition from $\alpha$-quartz to coesite occurs at 5~GPa, followed by
transformation to stishovite at $\sim$7~GPa, which continues to be stable up to
50~GPa. This phase transition sequence is in good agreement with experiments
\cite{Swamy1994} and with the GGA results by Demuth \emph{et al.}
\cite{Demuth1999}, Oganov \emph{et al.} \cite{Oganov2005} and LDA results of
Tsuchiya \emph{et al.} \cite{Tsuchiya2004}; it is known though \cite{Demuth1999}
that the GGA is more accurate than the LDA for phase transition pressures.
The new structure is not stable at any pressure for SiO$_{2}$, but at 0~GPa is
only 3.4~meV/atom higher in energy than $\alpha$-quartz, and should be
synthesizable as a metastable phase. Our results of coesite $\to$ coesite-II
transition are in good agreement with recent study of {\v Cernok} \emph{et al.}
\cite{Cernok2014}, where they observe coesite at 20.3~GPa, and after an abrupt
change in the diffraction pattern between $\sim$20 and $\sim$28~GPa ---
coesite-II at 27.5 and 30.9~GPa.

\subsubsection{Metastable structures of SiO$_{2}$}
It is well known that SiO$_{2}$ $\alpha$-quartz is thermodynamically stable at
ambient pressure. However, there are numerous known SiO$_{2}$ polymorphs which
are metastable, but exist in nature or can be synthesized. We examined SiO$_{2}$
feldspar, baddeleyite, melanophlogite and moganite at 0~GPa. El Goresy \emph{et
al.} \cite{Goresy2000} claimed a baddeleyite-like post-stishovite phase of
silica in the Shergotty meteorite, however later that controversial phase turned
out to be $\alpha$-PbO$_{2}$-like silica \cite{Dubrovinsky2001}. Our
calculations confirm that the baddeleyite-like form of SiO$_{2}$ is very
unfavorable at 0~GPa and spontaneously (barrierlessly) transforms into the
$\alpha$-PbO$_{2}$-like structure. We have found that SiO$_{2}$-feldspar,
moganite and melanophlogite are energetically very close to the stable phase
($\alpha$-quartz) and to the new $C2/c$ structure. Differences in enthalpy
between melanophlogite, new structure and $\alpha$-quartz are less than
0.01~eV/atom (see zoomed inset in Fig.~\ref{fig:enth_SiO2}). The fact that
complex open structure of melanophlogite (138~atoms/cell) has a slightly lower
energy than $\alpha$-quartz, can be explained by errors of the GGA, which were
discussed in details by Demuth \emph{et al.} \cite{Demuth1999}. They also found
$\beta$-cristobalite (Fig.~\ref{fig:enth_SiO2}) is lower in energy by about
0.03~eV/SiO$_{2}$ than $\alpha$-quartz, confirmed by calculations of Zhang
\emph{et al.} \cite{Zhang2014}, showing that the GGA slightly overstabilizes
low-density structures.

\subsection{Lattice dynamics}
\begin{figure}[htbp]
\includegraphics[scale=0.25]{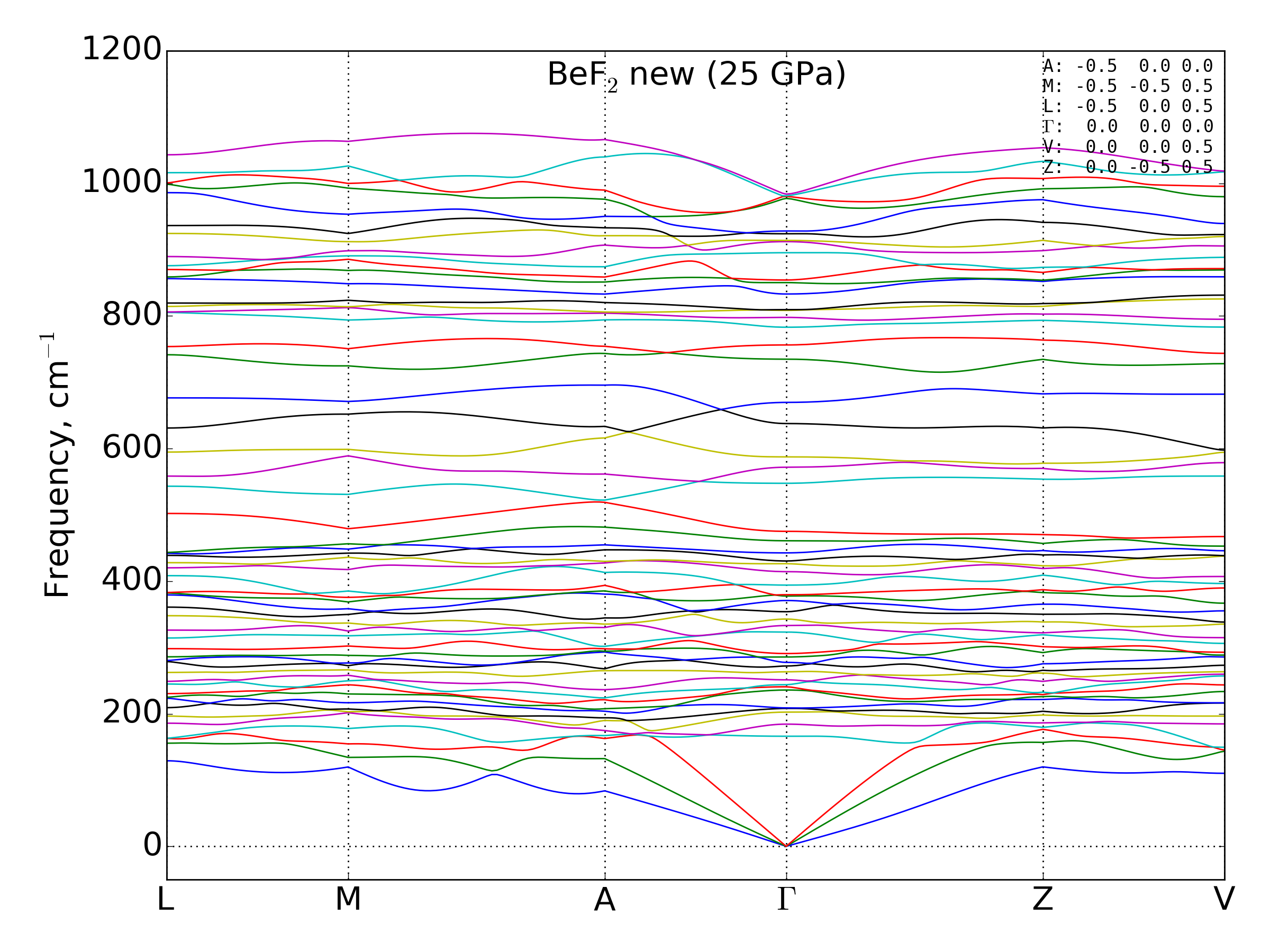}
\caption{\label{fig:phonons_BeF2_new_25GPa}Phonons dispersion curves showing
dynamical stability of the $C2/c$ structure of BeF$_{2}$ at 25~GPa.}
\end{figure}

Since the new structure of BeF$_{2}$ appears to be thermodynamically stable,
analysis of dynamical stability (phonon dispersion) has been performed for this
structure as well as for all other structures at pressures where they were found
to be thermodynamically stable. Our results show that BeF$_{2}$ $\alpha$-quartz
at 0~GPa, coesite at 5~GPa, new structure at 25~GPa and stishovite at 30~GPa do
not have imaginary frequencies. Similar results are observed for SiO$_{2}$
$\alpha$-quartz at 0~GPa, coesite at 5~GPa and stishovite at 10~GPa.
Fig.~\ref{fig:phonons_BeF2_new_25GPa} shows dynamical stability of the new
structure of BeF$_{2}$ since no imaginary frequencies are observed in the phonon
dispersion plot.

\subsection{Electronic properties}
\begin{figure}[htbp]

\includegraphics[scale=0.16]{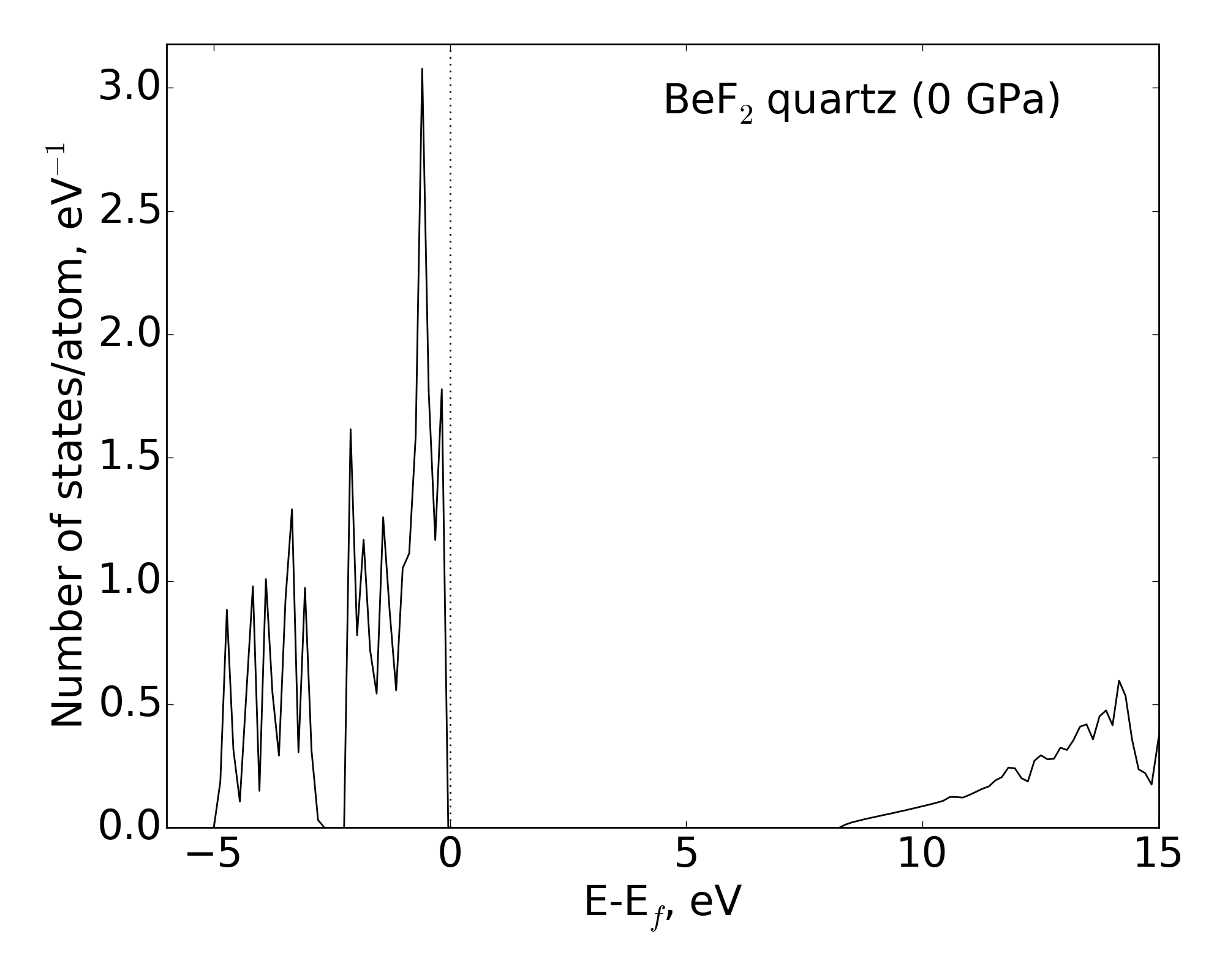}
\quad
\includegraphics[scale=0.16]{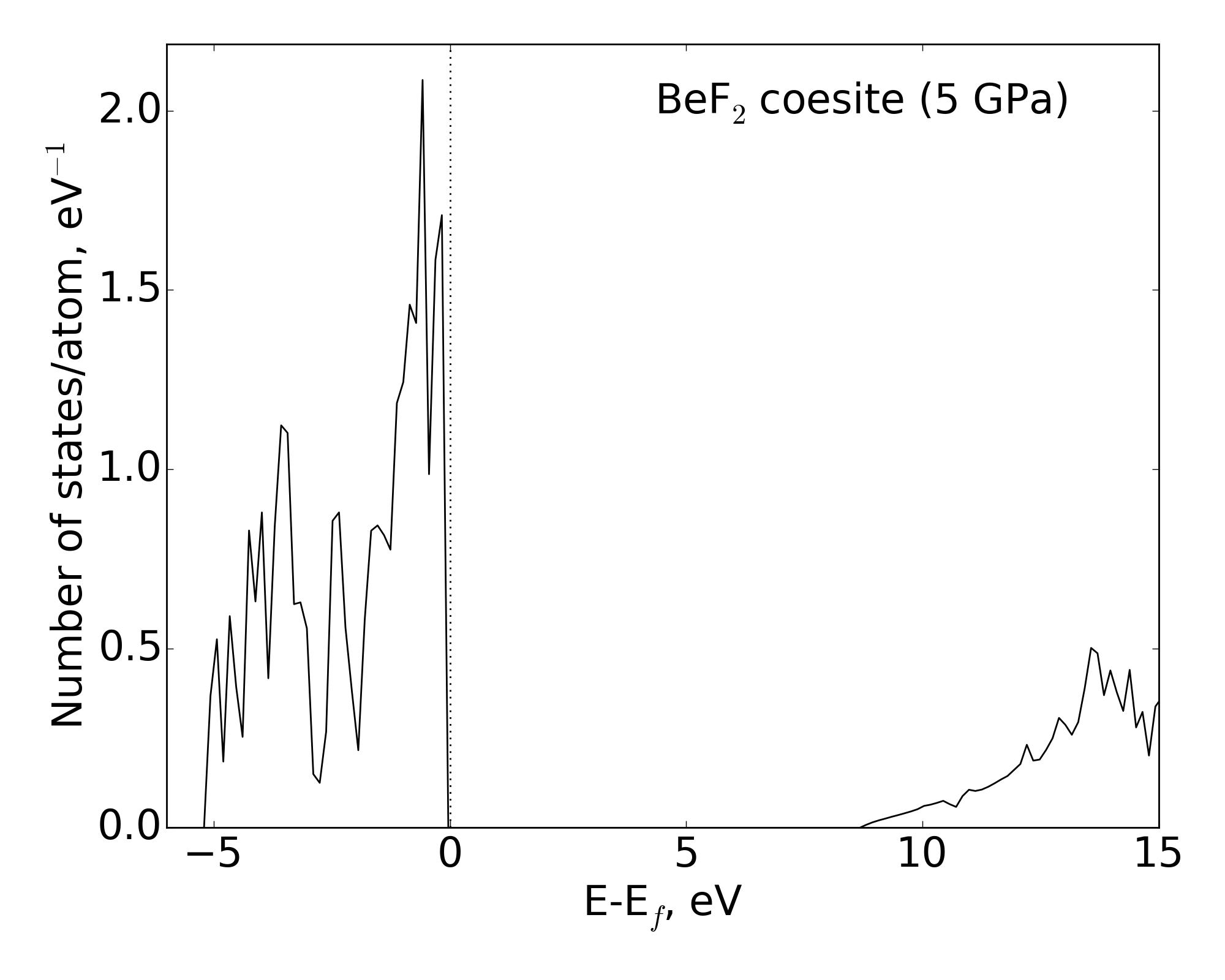}

\includegraphics[scale=0.16]{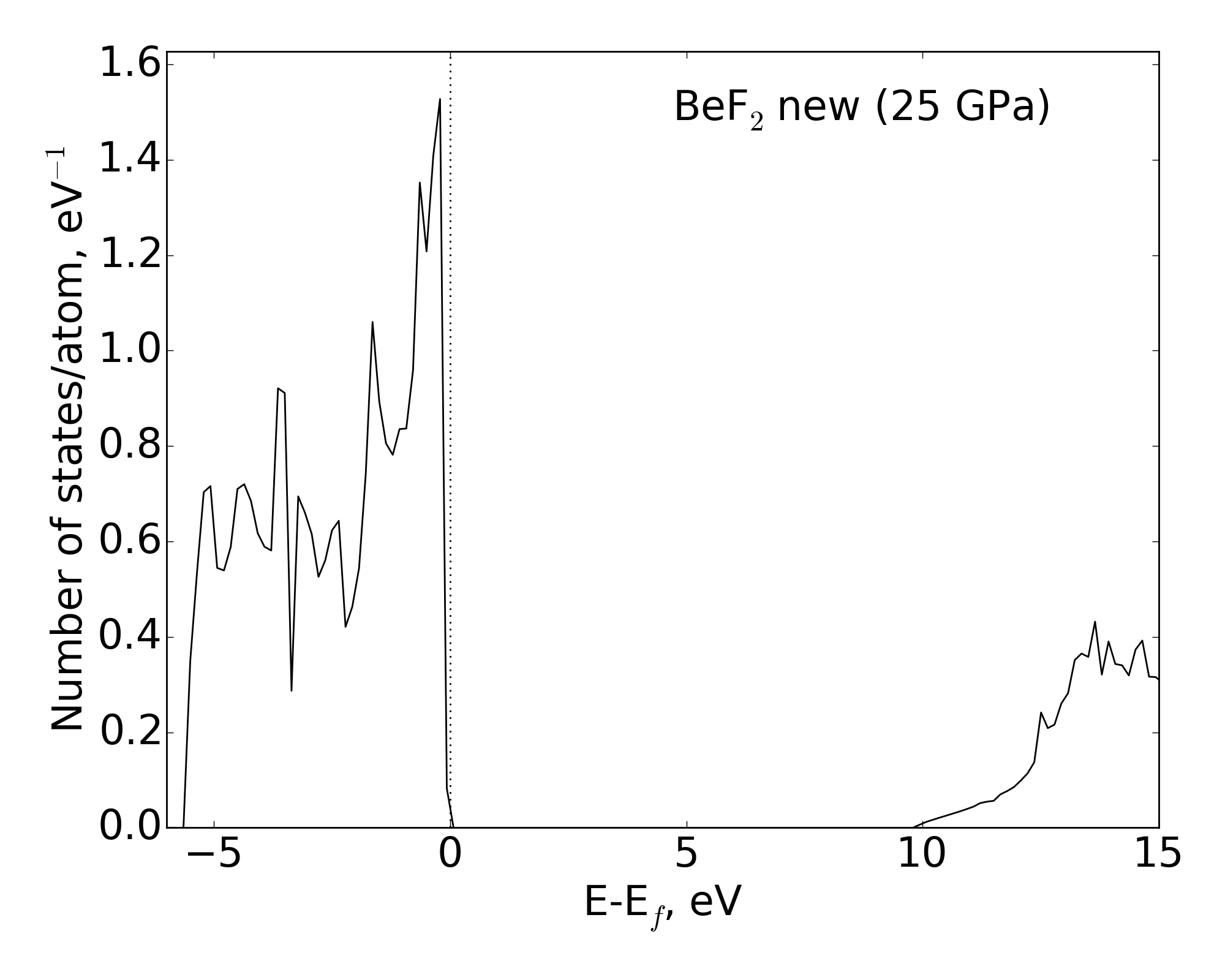}
\quad
\includegraphics[scale=0.16]{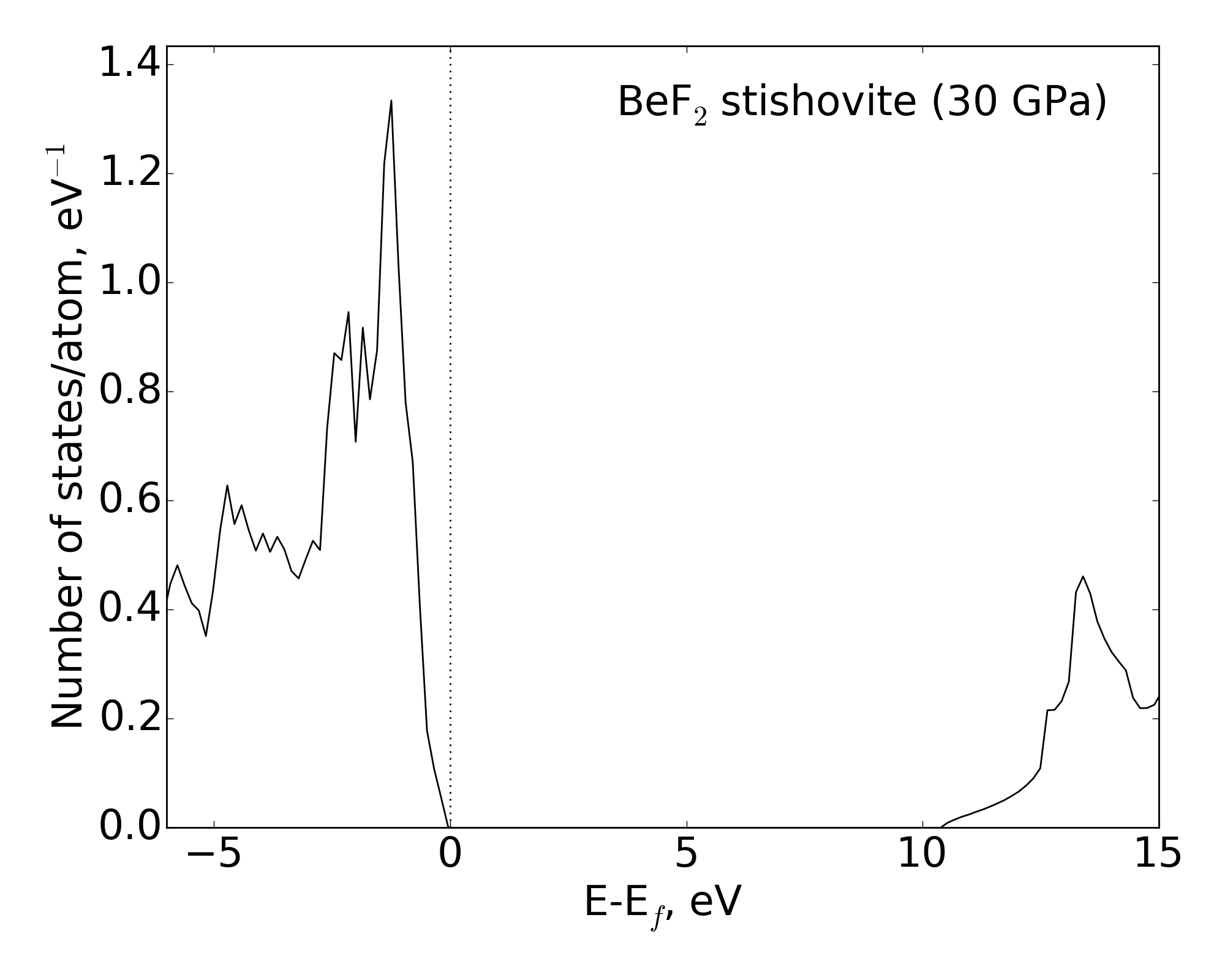}

\caption{\label{fig:DOS_BeF2}Density of states of BeF$_{2}$ in the
$\alpha$-quartz (at 0~GPa), coesite (at 5~GPa), $C2/c$ structure (at 25~GPa),
and stishovite (at 30~GPa) phases.}

\end{figure}

According to Fig.~\ref{fig:DOS_BeF2}, all BeF$_{2}$ phases are insulators, the
DFT band gap increases from $\sim$7 to $\sim$10~eV with increasing pressure from
0 to 30~GPa and the value of the gap is in good agreement with data of Yu
\emph{et al.} \cite{Yu2013}.

\begin{figure}[btp]

\subfigure[]{
\includegraphics[scale=0.25]{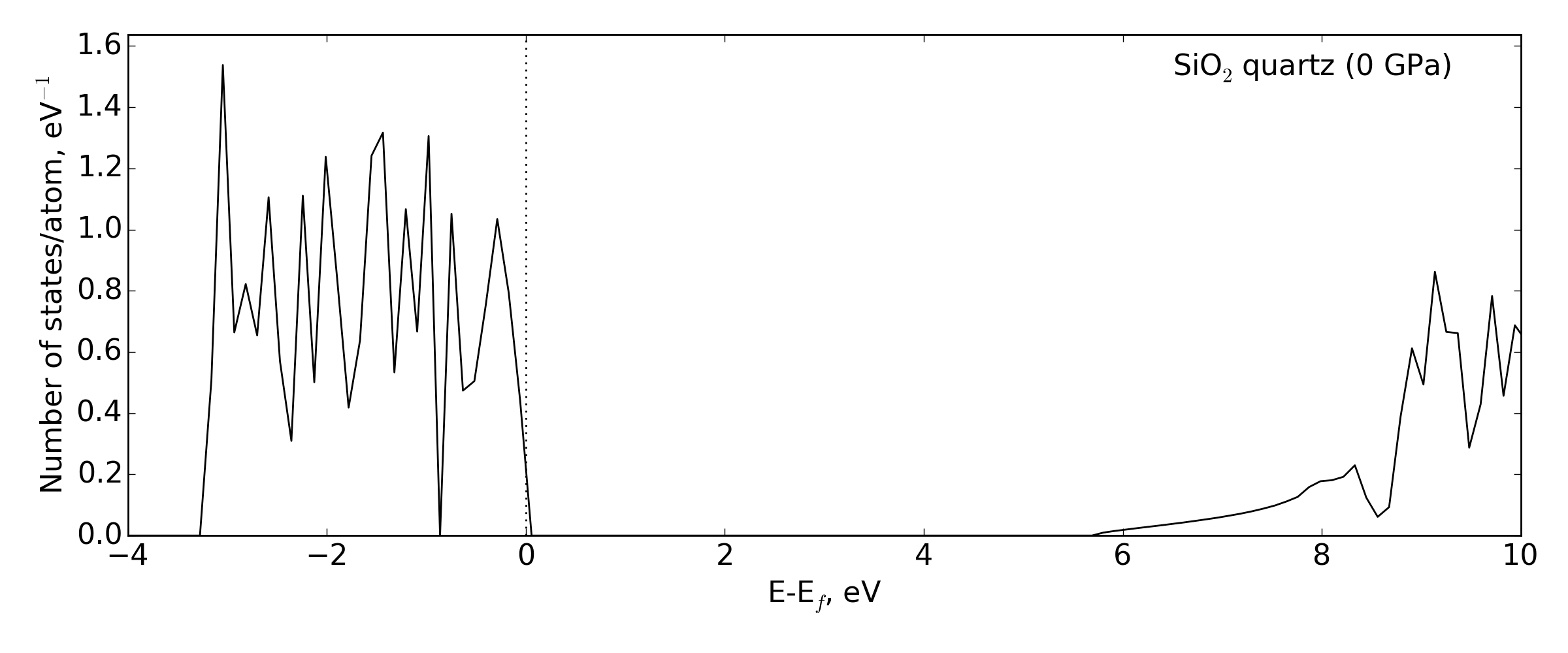}
\label{fig:DOS_SiO2_quartz}
}
\subfigure[]{
\includegraphics[scale=0.25]{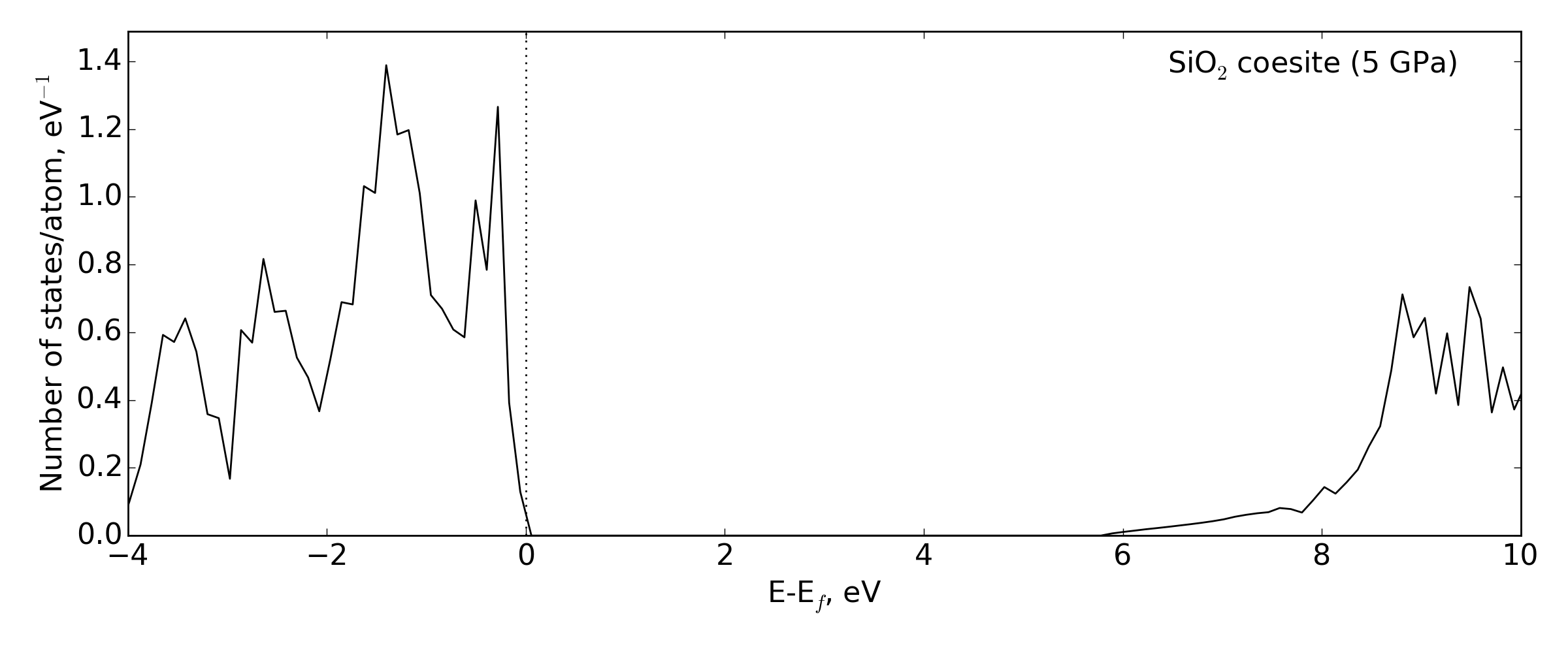}
\label{fig:DOS_SiO2_coesite}
}

\subfigure[]{
\includegraphics[scale=0.25]{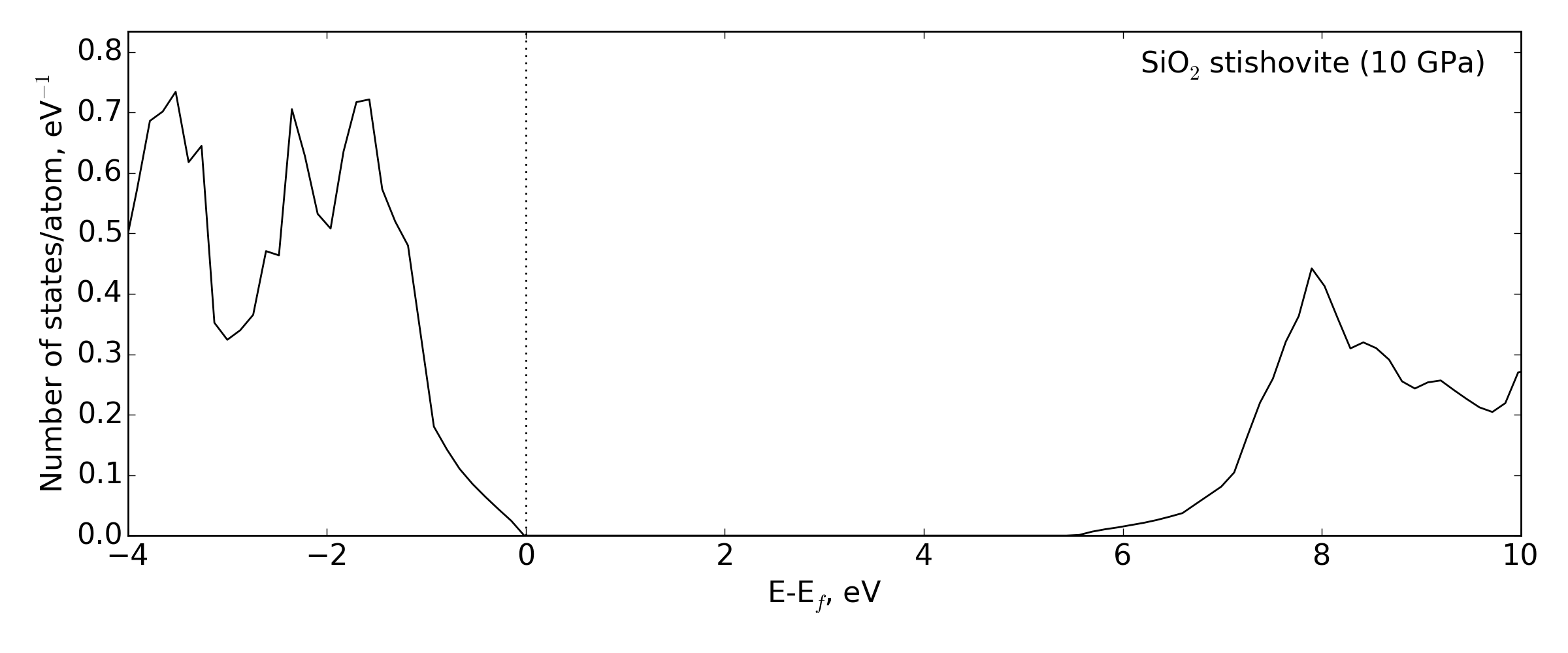}
\label{fig:DOS_SiO2_stishovite}
}

\caption{\label{fig:DOS_SiO2}Density of states of SiO$_{2}$ in the
\subref{fig:DOS_SiO2_quartz} $\alpha$-quartz (at 0~GPa),
\subref{fig:DOS_SiO2_coesite} coesite (at 5~GPa) and
\subref{fig:DOS_SiO2_stishovite} stishovite (at 10~GPa)
phases.
}

\end{figure}

For SiO$_{2}$ (Fig.~\ref{fig:DOS_SiO2}) we also observe insulating behavior, and
the band gap is about 6~eV and remains almost unchanged with increasing
pressure.

\subsection{Hardness}
Three models have been exploited to calculate hardnesses --- the Lyakhov-Oganov
\cite{Lyakhov2011}, Chen-Niu \cite{Chen2011} and Mukhanov-Kurakevych-Solozhenko
\cite{Mukhanov2008} models. First approach is based on concepts of bond
strengths and bond topology to compute hardness. Detailed description of the
methodology can be found in Ref. \cite{Lyakhov2011}. This model has been
implemented in the USPEX code, and for greater convenience has also been
implemented as an online utility available at
\url{http://han.ess.sunysb.edu/hardness/}. The second method of hardness
calculation is Chen-Niu model, which is based on elastic tensor components and
also implemented in the USPEX code. The third one is a thermodynamic model of
hardness.

The results can be seen in Table~\ref{tab:hardness}. Experimental data are
provided where available --- Vickers hardness of SiO$_{2}$-quartz
\cite{Oganov2010}, SiO$_{2}$-coesite \cite{Mukhanov2008} and
SiO$_{2}$-stishovite \cite{Leger1996}. From Table~\ref{tab:hardness} it is
clearly seen that the calculated hardness of SiO$_{2}$ quartz and stishovite is
much higher than one of BeF$_{2}$ analogs. The hardness of BeF$_{2}$ and
SiO$_{2}$ in the new $C2/c$ structure is comparable with the hardness of
$\alpha$-quartz and coesite.

\begin{table*}[htbp]
\caption{\label{tab:hardness}Hardness of BeF$_{2}$ and SiO$_{2}$ structures at
0~GPa in GPa. For the metastable SiO$_{2}$ structures we present enthalpies
relative to $\alpha$-quartz (in eV per formula unit).}
\begin{ruledtabular}
\begin{tabular}{ldddddd}
&
\multicolumn{3}{c}{\textrm{BeF$_{2}$}} &
\multicolumn{3}{c}{\textrm{SiO$_{2}$}} \\
& 
\multicolumn{1}{c}{\textrm{Lyakhov-Oganov}} &
\multicolumn{1}{c}{\textrm{Chen-Niu}}       &
\multicolumn{1}{c}{\textrm{Mukhanov \emph{et al.}}}\footnotemark[1] &
\multicolumn{1}{c}{\textrm{Lyakhov-Oganov}} &
\multicolumn{1}{c}{\textrm{Chen-Niu}}       &
\multicolumn{1}{c}{\textrm{Experiment}}     \\
\hline
Quartz         & 7.1 &  7.5 & 11.0 & 20.0 & 12.5 & 12.0\footnotemark[2]   \\
Coesite        & 8.2 &  8.3 & 11.7 & 22.3 &  8.4 & 20.0\footnotemark[2]   \\
New structure  & 7.3 &  6.8 & 13.5 & 19.1 &  6.7 & \multicolumn{1}{c}{\textrm{---}} \\
Stishovite     & 8.2 & 12.7 & 15.1 & 29.0 & 28.7 & 33.0\footnotemark[2]   \\
\hline
\multicolumn{7}{c}{\textrm{Metastable structures (SiO$_{2}$ only):}} \\
&
\multicolumn{2}{c}{\textrm{Relative enthalpy, eV/f.u.}} &
\multicolumn{4}{c}{\textrm{Hardness, GPa}}              \\
&
\multicolumn{2}{c}{}                              &
\multicolumn{2}{c}{\textrm{Lyakhov-Oganov model}} &
\multicolumn{2}{c}{\textrm{Chen-Niu model}}       \\
\hline
Feldspar       & \multicolumn{2}{d}{ 0.047} & \multicolumn{2}{d}{ 6.7} &
\multicolumn{2}{d}{11.8} \\
Baddeleyite    & \multicolumn{2}{d}{ 0.726} & \multicolumn{2}{d}{29.6} &
\multicolumn{2}{d}{28.0} \\
Melanophlogite & \multicolumn{2}{d}{-0.013} & \multicolumn{2}{d}{12.5} &
\multicolumn{2}{d}{ 3.3} \\
Moganite       & \multicolumn{2}{d}{ 0.003} & \multicolumn{2}{d}{19.5} &
\multicolumn{2}{d}{12.8} \\
\end{tabular}

\footnotetext[1]{Thermodynamic model of hardness (Ref.~\cite{Mukhanov2008})}
\footnotetext[2]{Vickers hardness}

\end{ruledtabular}
\end{table*}

\section{Conclusions}
We have examined thermodynamic, vibrational, electronic and elastic properties
of BeF$_{2}$ and SiO$_{2}$ phases using DFT calculations. The sequence of
pressure-induced phase transitions of BeF$_{2}$ up to 50~GPa is as follows:
$\alpha$-quartz-type $\xrightarrow{4~GPa}$ coesite-type $\xrightarrow{18~GPa}$
$C2/c$ $\xrightarrow{27~GPa}$ stishovite (rutile-type) structures. We found a
new phase of BeF$_{2}$ which is thermodynamically stable at pressures from 18 to
27~GPa. This phase is not observed in SiO$_{2}$, but could be synthesized in
principle. Electronic properties analysis has shown BeF$_{2}$ and SiO$_{2}$
remain insulating in a wide range of pressures (from 0 to 50~GPa). Hardness of
BeF$_{2}$ and SiO$_{2}$ in the new structure is comparable with hardness of
$\alpha$-quartz and coesite at 0~GPa. Hardnesses of metastable SiO$_{2}$
structures have been examined as well.

\section{Author contributions}
M.R., H.N. and M.D. performed the calculations, M.R. and A.R.O. contributed to
the analysis and wrote the paper. X.F.Z and G.R.Q. provided technical assistance
with calculations. V.L.S. proposed the idea, performed calculations of hardness
and participated in the discussion.

\section{Additional information}
Competing financial interests: The authors declare no competing financial
interests.

\begin{acknowledgments}
We thank the National Science Foundation (EAR-1114313, DMR-1231586), DARPA
(Grant No.~W31P4Q1210008), the Government of Russian Federation (grant
No.~14.A12.31.0003), and Foreign Talents Introduction and Academic Exchange
Program (No.~B08040). Also, we thank Dr. V.A.~Mukhanov for valuable comments.
\end{acknowledgments}

\appendix
\section{Densities of BeF$_{2}$ and SiO$_{2}$ structures}

Table~\ref{tab:density} shows densities of BeF$_{2}$ structures at 0 and
20~GPa and SiO$_{2}$ structures at 0~GPa.

\begin{table}[htbp]
\caption{\label{tab:density}Densities of BeF$_{2}$ and SiO$_{2}$ structures.}
\begin{ruledtabular}
\begin{tabular}{lddd}
\multicolumn{1}{c}{\textrm{System}}     &
\multicolumn{1}{c}{\textrm{Number of}}  &
\multicolumn{1}{c}{\textrm{Volume,}}    &
\multicolumn{1}{c}{\textrm{Density,}}   \\
&
\multicolumn{1}{c}{\textrm{atoms}}      &
\multicolumn{1}{c}{\textrm{\r{A}/cell}} &
\multicolumn{1}{c}{\textrm{g/cm$^{3}$}} \\
\hline
\multicolumn{4}{l}{\textrm{\textbf{BeF$_{2}$ at 0~GPa:}}} \\
$\alpha$-quartz         &  9 &  105.167 & 2.244 \\
coesite                 & 24 &  254.636 & 2.472 \\
coesite-II              & 96 & 1021.960 & 2.464 \\
$C2/c$                  & 18 &  213.696 & 2.209 \\
stishovite              &  6 &   47.771 & 3.294 \\
\hline
\multicolumn{4}{l}{\textrm{\textbf{BeF$_{2}$ at 20~GPa:}}} \\
$\alpha$-quartz         &  9 &   73.078 & 3.230 \\
coesite                 & 24 &  202.001 & 3.116 \\
$C2/c$                  & 18 &  145.159 & 3.252 \\
stishovite              &  6 &   41.492 & 3.793 \\
\hline
\multicolumn{4}{l}{\textrm{\textbf{SiO$_{2}$ at 0~GPa:}}} \\
$\alpha$-quartz         &  9 &  116.934 & 2.580 \\
coesite                 & 24 &  283.341 & 2.839 \\
coesite-II              & 96 & 1137.296 & 2.830 \\
$C2/c$                  & 18 &  243.569 & 2.477 \\
stishovite              &  6 &   48.185 & 4.174 \\
$\alpha$-PbO$_{2}$-type & 12 &   94.623 & 4.251 \\

\end{tabular}
\end{ruledtabular}
\end{table}

\section{CIF file of BeF$_{2}$ $C2/c$ structure at 20~GPa}
\small{
\begin{verbatim}
# CIF file
# This file was generated by FINDSYM (H.T. Stokes)

data_findsym-output
 
_symmetry_space_group_name_H-M 'C 1 2/c 1'
_symmetry_Int_Tables_number 15
 
_cell_length_a       8.74241
_cell_length_b       8.69478
_cell_length_c       4.17800
_cell_angle_alpha   90.00000
_cell_angle_beta    66.07927
_cell_angle_gamma   90.00000
 
loop_
_space_group_symop_operation_xyz
x,y,z
-x,y,-z+1/2
-x,-y,-z
x,-y,z+1/2
x+1/2,y+1/2,z
-x+1/2,y+1/2,-z+1/2
-x+1/2,-y+1/2,-z
x+1/2,-y+1/2,z+1/2
 
loop_
_atom_site_label
_atom_site_type_symbol
_atom_site_fract_x
_atom_site_fract_y
_atom_site_fract_z
_atom_site_occupancy
Be1 Be   0.30175   0.08755   0.27471   1.00000
Be2 Be   0.00000   0.18404   0.25000   1.00000
F1  F   -0.11350   0.09592   0.11433   1.00000
F2  F    0.14707   0.43232   0.42082   1.00000
F3  F   -0.11681   0.27080  -0.42582   1.00000
\end{verbatim}
}


\bibliography{BeF2_article} 

\end{document}